\documentclass[journal]{IEEEtran}

\usepackage{amsmath}
\usepackage{amssymb}
\usepackage{amsfonts}
\usepackage{amsthm}
\usepackage{cite}
\usepackage{algorithm}
\usepackage{algpseudocode}
\usepackage{float}
\usepackage{tikz}

\newtheorem{theorem}{Theorem}
\newtheorem{lemma}{Lemma}
\newtheorem{corollary}{Corollary}
\newtheorem{definition}{Definition}

\newcommand{\blockcomment}[1]{}
\newcommand{\wrap}[1]{\left(#1\right)}
\newcommand{\stack}[2]{%
  \begin{tabular}{l}
    #1\\
    #2\\
  \end{tabular}}

\newcommand{\rstack}[2]{%
  \begin{tabular}{r}
    #1\\
    #2\\
  \end{tabular}}

\newcommand{\con}{\mathbin{:}}

\begin{document}
\title{An Improvement to Levenshtein's Upper Bound on the Cardinality of Deletion Correcting Codes}
\author{Daniel~Cullina,~\IEEEmembership{Student Member,~IEEE}
        and Negar~Kiyavash,~\IEEEmembership{Member,~IEEE}%
\thanks{The material in this paper was presented (in part) at the International Symposium on Information Theory, Istanbul, Turkey, July 2013~\cite{cullina_improvement_2013}.
This work was supported in part by NSF grants CCF 10-54937 CAR and CCF 10-65022 Kiyavash.}%
\thanks{Daniel Cullina is with the Department of Electrical and Computer Engineering and the Coordinated Science Laboratory, University of Illinois at Champaign-Urbana, Urbana, Illinois 61801 (email: cullina@illinois.edu). }%
\thanks{Negar Kiyavash is with the Department of Industrial and Enterprise Systems Engineering and the Coordinated Science Laboratory, University of Illinois at Champaign-Urbana, Urbana, Illinois 61801 (email: kiyavash@illinois.edu). }%
}
\maketitle
\begin{abstract}
We consider deletion correcting codes over a $q$-ary alphabet.
It is well known that any code capable of correcting $s$ deletions can also correct any combination of $s$ total insertions and deletions.
To obtain asymptotic upper bounds on code size, we apply a packing argument to channels that perform different mixtures of insertions and deletions.
Even though the set of codes is identical for all of these channels, the bounds that we obtain vary.
Prior to this work, only the bounds corresponding to the all insertion case and the all deletion case were known.
We recover these as special cases.
The bound from the all deletion case, due to Levenshtein, has been the best known for more than forty five years.
Our generalized bound is better than Levenshtein's bound whenever the number of deletions to be corrected is larger than the alphabet size.
\end{abstract}

\section{Introduction}
\IEEEPARstart{D}{eletion} channels output only a subsequence of their input while preserving the order of the transmitted symbols. 
Deletion channels are related to synchronization problems, a wide variety of problems in bioinformatics, and the communication of information over packet networks.
This paper concerns channels that take a fixed length input string of symbols drawn from a $q$-ary alphabet and delete a fixed number of symbols. 
In particular, we are interested in upper bounds on the cardinality of the largest possible $s$-deletion correcting codebook.

Levenshtein derived asymptotic upper and lower bounds on the sizes of binary codes for any number of deletions~\cite{levenshtein_binary_1966}.
These bounds easily generalize to the $q$-ary case~\cite{tenengolts_nonbinary_1984}.
He showed that the Varshamov Tenengolts (VT) codes, which had been designed to correct a single asymmetric error~\cite{varshamov_codes_1965,varshamov_arithmetic_1965}, could be used to correct a single deletion.
The VT codes establish the asymptotic tightness of the upper bound in the case of a binary alphabet and a single deletion.

Since then, a wide variety of code constructions, which provide lower bounds, have been proposed for the deletion channel and other closely related channels.
One recent construction uses constant Hamming weight deletion constructing codes~\cite{cullina_coloring_2012}.
In contrast, progress on upper bounds has been rare.
Levenshtein eventually refined his original asymptotic bound (and the parallel nonbinary bound of Tenengolts) into a nonasymptotic version~\cite{levenshtein_bounds_2002}.
Kulkarni and Kiyavash recently proved a better upper bound for an arbitrary number of deletions and any alphabet size~\cite{kulkarni_non-asymptotic_2012}.

Another line of work has attacked some related combinatorial problems.
These include characterization of the sets of superstrings and substrings of any string.
Levenshtein showed that the number of superstrings does not depend on the starting string~\cite{levenshtein_elements_1974}.
He also gave upper and lower bounds on the number of substrings using the number of runs in the starting string~\cite{levenshtein_binary_1966}.
Calabi and Hartnett gave a tight bound on the number of substrings of each length~\cite{calabi_general_1969}.
Hirschberg extended the bound to larger alphabets~\cite{hirschberg_bounds_1999}.
Swart and Ferreira gave a formula for the number of distinct substrings produced by two deletions for any starting string~\cite{swart_note_2003}.
Mercier et al showed how to generate corresponding formulas for more deletions and gave an efficient algorithm to count the distinct substrings of any length of a string~\cite{mercier_number_2008}.
Liron and Langberg improved and unified existing bounds and constructed tightness examples~\cite{liron_characterization_2012}.
Some of our intermediate results contribute to this area.

\subsection{Upper bound technique}
\label{section:motivation}
To derive our upper bounds, we use a packing argument that can be applied to any combinatorial channel.
Any combinatorial channel can be represented by a bipartite graph.
Channel inputs correspond to left vertices, channel outputs correspond to right vertices, and each edge connects an input to an output that can be produced from it.
If two channel inputs share a common output, they cannot both appear in the same code.
The degree of an input vertex in the graph is the number of possible channel outputs for that input.
If the degree of each input is at least $r$ and there are $N$ possible outputs, any code contains at most $N/r$ codewords.
For a channel that makes at most $s$ substitution errors, this argument leads to the well known Hamming bound.

Any code capable of correcting $s$ deletions is also capable of correcting any combination of $s$ total insertions and deletions (See Lemma~\ref{lemma:channel-equiv}).
Despite this equivalence, this packing argument produces different upper bounds for channel that perform different mixtures of insertions and deletions.
Let $C_{q,s,n}$ be the size of the largest $q$-ary $n$-symbol $s$-deletion correcting code.
Prior to this work, the bounds on $C_{q,s,n}$ coming from the $s$-insertion channel and the $s$-deletion channel were known.

For the $s$-insertion channel, each $q$-ary $n$-symbol input has the same degree.
For fixed $q$ and $s$, the degree is asymptotic to $\binom{n}{s}(q-1)^s$ (See \eqref{eq:insertion-number}).
There are $q^{n+s}$ possible outputs, so 
\begin{equation}
\label{eq:bad-upper}
C_{q,s,n} \lesssim \frac{q^{n+s}}{\binom{n}{s}(q-1)^s}.
\end{equation}

The $s$-deletion case is slightly more complicated because different inputs have different degrees.
For instance, the input strings consisting of a single symbol repeated $n$ times have only a single possible output: the string with that symbol repeated $n-s$ time.
Consequently, using the minimum degree over all of the inputs yields a worthless bound.
Using the following argument~\cite{levenshtein_binary_1966}, Levenshtein showed that 
\begin{equation}
\label{eq:lev-upper}
C_{q,s,n} \lesssim \frac{q^n}{\binom{n}{s}(q-1)^s}.
\end{equation}
The average degree of an input is asymptotic to $\left(\frac{q-1}{q}\right)^s\binom{n}{s}$ and most inputs have a degree close to that.
The inputs can be divided into two classes: those with degree at least $1-\epsilon$ times the average degree and those with smaller degree. 
For an appropriately chosen $\epsilon$ that goes to zero as $n$ goes to infinity, the vast majority of inputs fall into the former class.
Call members of the former class the typical inputs.
The minimum degree argument can be applied to bound the number of typical inputs that can appear in a code.
There are $q^{n-s}$ possible outputs, so the number of typical inputs in a code is asymptotically at most \eqref{eq:lev-upper}.
We have no information about what the fraction of the atypical inputs can appear in a code, but the total number of atypical inputs is small enough to not affect the asymptotics of the upper bound.

The bounds \eqref{eq:bad-upper} and \eqref{eq:lev-upper} have the same growth rates, but the bound on deletion correcting codes is a factor of $q^s$ better than the bound on insertion correcting codes, despite the fact that any $s$-deletion correcting code is an $s$-insertion correcting code and vice versa.
Note that there is no possible improvement to the insertion channel bound from dividing the inputs into typical and atypical classes.  

We extend this bounding strategy to channels that perform both deletions and insertions.
We obtain a generalized upper bound that includes Levenshtein's bound as a special case.
Recall that Levenshtein's bound is known to be tight for one deletion and alphabet size two.
The new bound improves upon the Levenshtein's bound whenever the number of deletions is greater than the alphabet size.

The rest of the paper is organized as follows.
In Section~\ref{section:preliminaries}, we present some notation and basic results on deletion and insertion channels.
In Section~\ref{section:edges}, we construct a class of well-behaved edges in the channel graph.
Together with an upper bound on the number of edges in the channel graph, the size of this class establishes the asymptotics of the average input degree.
In Section~\ref{section:bounds}, we prove a lower bound on the degree of each input vertex and use it to establish our main result: an upper bound on the size of a $q$-ary $s$-deletion correcting code.

\section{Preliminaries}
\label{section:preliminaries}
\subsection{Notation}
Let $\mathbb{N}$ be the set of nonnegative integers.
Let $[n]$ be the set of nonnegative integers less than $n$, $\{0,1,\dots,n-1\}$.
Let $[q]^n$ be the set of $q$-ary strings of length $n$.
Let $[q]^*$ be the set of $q$-ary strings of all lengths.
More generally, for a set $S$, let $S^n$ be the set of lists of elements $S$ of length $n$ and let $S^*$ be the set of lists of elements of $S$ of any length.

We will need the following asymptotic notation: 
let $a(n) \sim b(n)$ denote that $\lim_{n \to \infty} \frac{a(n)}{b(n)} = 1$
and $a(n) \lesssim b(n)$ denote that $\lim_{n \to \infty} \frac{a(n)}{b(n)} \leq 1$.
We will use the following asymptotic equality frequently: for fixed $c$, $\binom{n}{c} \sim \frac{n^c}{c!}$.

\subsection{Deletion distance}
The substring relation is a partial ordering of $[q]^*$.
Consequently for strings $x$ and $y$, we write $x \preceq y$ if $x$ is a substring of $y$.
\begin{definition}
For $x \in [q]^n$ and $y \in [q]^m$, define the deletion distance between them to be $d_L(x,y) = n + m - 2l$, where $l$ is the length of their longest common substring.
\end{definition}
It is well known that deletion distance is a metric.
We will need a slightly stronger property.
The following lemma is the source of the nice properties of the deletion distance.
\begin{lemma}
\label{lemma:parallelogram}
For $l,m,n \in \mathbb{N}$ with $l \leq m$ and $l \leq n$, let $x \in [q]^n$ and $y \in [q]^m$.
Then there exists $z \in [q]^l$ such that $x \succeq z$ and $y \succeq z$ if and only if there exists $w \in [q]^{m+n-l}$ such that $w \succeq x$ and $w \succeq y$.
\end{lemma}
\begin{IEEEproof}
Given $x$, $y$, and $w$, a canonical $z$ can be constructed by a simple greedy algorithm.
Given $x$, $y$, and $z$, at least one $w$ can be constructed by a similar algorithm.
\end{IEEEproof}

The next lemma is a strengthening of the triangle inequality.
\begin{lemma}
\label{lemma:all-lengths}
For $l,m,n \in \mathbb{N}$ with $l \leq m$ and $l \leq n$, let $a=n-l$ and $b=m-l$.
For $x \in [q]^n$ and $y \in [q]^m$, the following are equivalent:
\begin{itemize}
\item[$A$] There exists $z \in [q]^*$ such that $d_L(x,z) \leq a$ and $d_L(y,z) \leq b$.
\item[$B$] $d_L(x,y) \leq a+b$
\item[$C$] For all $0 \leq i \leq a+b$, there exists $z_i \in [q]^{l+2i}$ such that $d_L(x,z_i) \leq a$ and $d_L(y,z_i) \leq b$.
\end{itemize}
\end{lemma}
\begin{IEEEproof}
$(A \Rightarrow B)$ Let the length of $z$ be $k$. 
Because $d(x,z)=a$, $x$ and $z$ have a common substring $u$ of length $(n+k-a)/2$. 
Similarly $y$ and $z$ have a common substring $v$ of length $(m+k-b)/2$.
By Lemma~\ref{lemma:parallelogram}, $u$ and $v$ have a common substring $w$ of length $(n+k-a)/2 + (m+k-b)/2 - k = (m+n-a-b)/2 = l$.
Because $w$ is a substring of both $x$ and $y$, $d(x,y) \leq a + b$.

$(B \Rightarrow C$) Let $z_0$ be a common substring of $x$ and $y$ of length $l$. 
There are $u_i,v_i \in [q]^{l+i}$ such that $x \succeq u_i \succeq z_0$ and $y \succeq v_i \succeq z_0$.
By Lemma~\ref{lemma:parallelogram}, $u_i$ and $v_i$ have a common superstring $z_i$ of length $2(l+i)-l = l+2i$.
Because $u_i$ is a common substring of $x$ and $z_i$, $d(x,z_i) \leq a$.
Similarly $d(y,z_i) \leq b$.

$(C \Rightarrow A$) Trivial.
\end{IEEEproof}

\begin{corollary}
\label{lemma:metric}
Deletion distance is a metric.
\end{corollary}
\begin{IEEEproof}
Deletion distance is symmetric.
Because $x$ is a substring of itself, $d(x,x)=0$.
Because the only substring of $x$ with the same length is $x$, $d(x,y)=0$ implies $x=y$.
From Lemma~\ref{lemma:all-lengths}, deletion distance satisfies the triangle inequality.
\end{IEEEproof}

\subsection{Deletion and insertion channels}
\label{section:deletion-channel}
We formalize the problem of correcting deletions and insertions by defining the following sets.
\begin{definition}
For $x \in [q]^n$, define $S_{a,0}(x) = \{z~\in~[q]^{n-a}:z~\preceq~x\}$, 
the set of substrings of $x$ that can be produced by $a$ deletions.
Define $S_{0,b}(x) = \{w \in [q]^{n+b} : w \succeq x \}$,
the set of superstrings of $x$ that can be produced by $b$ insertions.
Define $S_{a,b}(x) = \bigcup_{z \in S_{a,0}(x)} S_{0,b}(z)$.
\end{definition}
The $a$-deletion $b$-insertion channel takes a string of length $n$, finds a substring of length $n-a$, and outputs a superstring of that substring of length $n-a+b$.
Consequently, for each input $x$ to an $n$-symbol $a$-deletion $b$-insertion channel $S_{a,b}(x)$ is the set of possible outputs.

The following graph completely describes the behavior of the $(l+a)$-symbol $a$-deletion $b$-insertion channel.
\begin{definition}
Let $B_{q,l,a,b}$ be a bipartite graph with left vertex set $[q]^{l+a}$ and right vertex set $[q]^{l+b}$.
Vertices are adjacent if they have a common substring of length $l$.
\end{definition}
If $x$ is a left vertex of $B_{q,l,a,b}$, then its neighborhood is $S_{a,b}(x)$.
When two inputs share common outputs they can potentially be confused by the receiver.
\begin{definition}
A $q$-ary $n$-symbol $a$-deletion $b$-insertion correcting code is a set $C \subset [q]^n$ such that for any two distinct strings $x,y \in C$, $S_{a,b}(x) \cap S_{a,b}(y)$ is empty.
\end{definition}

\begin{lemma}
\label{lemma:channel-equiv}
For $a,b,n \in \mathbb{N}$, $x,y \in [q]^n$, $S_{a,b}(x) \cap S_{a,b}(y) = \varnothing$ if and only if $d_L(x,y) > 2(a+b)$.
Consequently a set $C \subset [q]^n$ is a $q$-ary $n$-symbol $a$-deletion $b$-insertion correcting code if and only if for all distinct $x,y \in C$, $d_L(x,y) > 2(a+b)$.
\end{lemma}
\begin{IEEEproof}
Let $s=a+b$.
Suppose there is some $z \in S_{a,b}(x) \cap S_{a,b}(y)$. 
Then $d_L(x,z) \leq s$ and $d_L(y,z) \leq s$, so $d(x,y)~\leq~2s$.

If $d_L(x,y) \leq 2s$, then by Lemma~\ref{lemma:all-lengths} there is some $w \in [q]^{n-a+b}$ such that $d_L(x,w) \leq s$ and $d_L(y,w) \leq s$.
\end{IEEEproof}

\section{Constructing edges}
\label{section:edges}
To execute the strategy described in section~\ref{section:motivation}, we need a lower bound on the degree of each channel input.
This is a lower bound on the degree of each left vertex of $B_{q,l,a,b}$.
To obtain this bound, we first construct a subset of the edges of $B_{q,l,a,b}$ that is easier to work with than the complete edge set.
Our ultimate lower bound on the degree of an input will actually be a lower bound on the number of edges for this subset incident to the input vertex.

One way to get information about the size of a target set $T$ is to find a construction function $f:P \rightarrow T$, where $P$ is an easily counted parameter set.
If $f$ is injective, then $|P| = |f(P)|$ and $|P| \leq |T|$.
We can demonstrate the injectivity of $f$ with a deconstruction function $g:T \rightarrow P$ that is a left inverse of $f$.
This means that $g(f(p)) = p$ for all $p \in P$.
If the function $g$ is given a constructible member of $T$, $g$ recovers the construction parameters that produce it.
Similarly, if $f$ is surjective, then we can find an injective $g:T \rightarrow P$ that is a right inverse of $f$, so $|T| = |g(T)|$ and $|P| \geq |T|$.
If $f$ is both injective and surjective, then $|P| = |T|$.

In this section we apply this method to the edge set of $B_{q,l,a,b}$.
We give an upper bound on the number of edges and briefly discuss why it is difficult to count the edges exactly.
We explain our construction of a subset of the edges and prove a lower bound on the size of this subset.
Finally we show that the upper and lower bounds match asymptotically.

\subsection{An upper bound}
By definition, two vertices in $B_{q,l,a,b}$ are adjacent if they share a substring of length $l$.
This makes the common substring a natural construction parameter for the edge.
We can construct an edge by starting with a string of length $l$, performing $a$ arbitrary insertions to obtain the left vertex, and performing $b$ arbitrary insertions to obtain the right vertex.
Our upper bound will use the following fact about insertions due to Levenshtein~\cite{levenshtein_elements_1974}.
 Each $x \in [q]^{n-s}$ has the same number of superstrings of length $n$:
\begin{equation}
|S_{0,s}(x)| = I_{q,s,n}, 
\label{eq:insertion-number}
\end{equation}
where
\begin{equation*}
I_{q,s,n} = \sum_{i=0}^s \binom{n}{i}(q-1)^i.
\end{equation*}
For fixed $q$ and $s$, $I_{q,s,n} \sim \binom{n}{s}(q-1)^s$.

\begin{lemma}
\label{lemma:edge-ub}
For all $q,l,a,b \in \mathbb{N}$ with $s=a+b$, the number of edges in $B_{q,l,a,b}$ satisfies 
\begin{IEEEeqnarray*}{rCl}
|E(B_{q,l,a,b})| &\leq& q^l I_{q,a,l+a} I_{q,b,l+b}\\
&\sim& q^l\binom{l}{a}(q-1)^a\binom{l}{b}(q-1)^b\\
&\sim& q^l\binom{l}{s}\binom{s}{a}(q-1)^{s}.
\end{IEEEeqnarray*}
\end{lemma}
\begin{IEEEproof}
There are $q^l I_{q,a,l+a} I_{q,b,l+b}$ triples $(z,x,y) \in [q]^l \times [q]^{l+a} \times [q]^{l+b}$ such that $z \preceq x$ and $z \preceq y$.
If $x \in [q]^{l+a}$ and $y \in [q]^{l+b}$ are adjacent in $B_{q,l,a,b}$, then they have at least one common substring of length $l$ and appear in at least one triple.
\end{IEEEproof}

This upper bound is not an equality because many pairs of strings $(x,y) \in [q]^{l+a} \times [q]^{l+b}$ have multiple common substrings $z \in [q]^l$.
Pairs of strings with multiple common substrings of length $l$ fall into two classes.
Pairs in the first class have a common substring of length more than $l$.
Call this string $w$.
In this case, every substring of length $l$ of $w$ is a common substring of the pair.
Pairs in the second class have multiple maximum length common substrings.
For example, the strings $0101$ and $1010$ have both $010$ and $101$ as substrings.

To determine the exact number of edges in $B_{q,l,a,b}$, it is necessary to determine the sizes of both classes.
The size of the first class can be found easily if the number of edges in $B_{q,l+i,a-i,b-i}$ is known for all $i$ up to $\min(a,b)$.
It is more difficult to characterize the vertex pairs of the second class.
Consequently, our lower bound will also not be tight.

\subsection{Constructing edges at most once each}
\label{subsection:construct}
\begin{figure}
\centering
 \begin{tikzpicture}[
 scale = 1/3,
 text height = 1.5ex,
 text depth =.1ex,
 b/.style={very thick}]

 \draw[b]  (8, 1) --  (2, 6);
 \draw[b]  (8, 1) --  (3, 6);
 \draw[b] (16, 1) -- (11, 6);
 \draw[b] (16, 1) -- (12, 6);

 \draw[b] (11, 1) -- (16, 3);
 \draw[b] (11, 1) -- (17, 3);

 \draw[b]  (0, 6) rectangle  (2, 7);
 \draw[b]  (2, 6) rectangle  (3, 7);
 \draw[b]  (3, 6) rectangle  (6, 7);
 \draw[b]  (6, 6) rectangle (11, 7);
 \draw[b] (11, 6) rectangle (12, 7);
 \draw[b] (12, 6) rectangle (15, 7);

 \draw[fill=white] (11, 3) rectangle (25, 4);
 \draw[b] (11, 3) rectangle (13, 4);
 \draw[b] (13, 3) rectangle (16, 4);
 \draw[b] (16, 3) rectangle (17, 4);
 \draw[b] (17, 3) rectangle (22, 4);
 \draw[b] (22, 3) rectangle (25, 4);

 \draw[b] (6, 0) rectangle (8, 1);
 \draw[b] (8, 0) rectangle (11, 1);
 \draw[b] (11, 0) rectangle (16, 1);
 \draw[b] (16, 0) rectangle (19, 1);

 \foreach \i in {3,10}
    \draw (\i + 6.5, 0.5) node {0};
 \foreach \i in {0,1,2,6,8,9,12}
    \draw (\i + 6.5, 0.5) node {1};
 \foreach \i in {4,5,7,11}
    \draw (\i + 6.5, 0.5) node {2};

 \foreach \i in {4,12}
    \draw (\i + 0.5, 6.5) node {0};
 \foreach \i in {0,1,3,7,9,10,14}
    \draw (\i + 0.5, 6.5) node {1};
 \foreach \i in {2,5,6,8,11,13}
    \draw (\i + 0.5, 6.5) node {2};

 \foreach \i in {3,11}
    \draw (\i + 11.5, 3.5) node {0};
 \foreach \i in {0,1,2,5,7,9,10,13}
    \draw (\i + 11.5, 3.5) node {1};
 \foreach \i in {4,6,8,12}
    \draw (\i + 11.5, 3.5) node {2};

 \draw (7.5,8) node {$x$};
 \draw (18, 5) node {$y$};
 \draw (12.5, -1) node {$z$};

 \end{tikzpicture}
\caption{An example of an edge $(x,y) \in E(B_{3,13,2,1})$ constructed from a common substring $z\in[3]^{13}$.}
\label{figure:strings}
\end{figure}
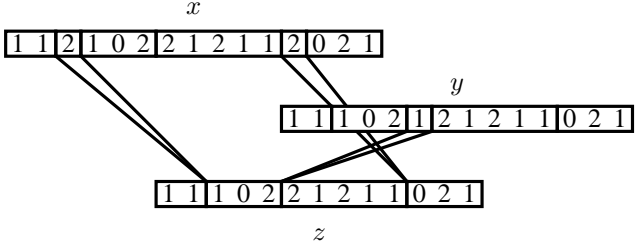

Our lower bound uses a different construction.
To construct an edge $(x,y) \in E(B_{q,l,a,b})$, start with a string $z \in [q]^l$.
As before, $z$ will be a substring of both endpoints of the edge.
Let $s = a+b$.
Partition $z$ into $s+1$ nonempty intervals.
To produce $x$, select $a$ of the $s$ boundaries between intervals and insert one new symbol into $z$ at each.
To produce $y$, insert one new symbol into $z$ at each of the other $b$ boundaries.
Figure~\ref{figure:strings} gives an example.

Each way to partition $z$ corresponds to a composition of $l$ with $s+1$ parts.
\begin{definition}
A composition of $l$ with $t$ parts is a list of $t$ nonnegative integers with sum $l$.
Let $M(t,l,k)$ be the family of compositions of $l$ with $t$ parts and each part of size at least $k$:
\begin{equation*}
M(t,l,k)~=~\left\{\lambda \in (\mathbb{N} \setminus [k])^t \left| \sum_{i \in [t]} \lambda_i = l
\right.\right\}.
\end{equation*}
\end{definition}
A standard argument shows that $|M(t,l,k)| = \binom{l-kt+t-1}{t-1}$.

Thus the parameter set for this construction is
\begin{equation*}
[q]^l \times M(s+1,l,1) \times \binom{[s]}{a} \times [q]^s
\end{equation*}
where $\binom{[s]}{a}$ is the family of $a$ element subsets of $[s]$.
The size of this set is $\binom{l-1}{s}\binom{s}{a}q^{l+s}$.

It is clear that there are many edges that this construction produces multiple times.
We will show that if the following two restrictions are added to construction procedure, each edge will be produced at most once:
\begin{itemize}
\item Each inserted symbol must differ from the leftmost symbol in the interval to its right.
\item Each interval of $z$ must be nonalternating.
\end{itemize}

The first restriction is well posed because the intervals are nonempty.
This restriction is needed because inserting a new symbol anywhere within a run of that same symbol has the same effect.
Under the restriction, a run in $z$ can only be extended by inserting a matching symbol at the right end.
To implement this restriction, for each insertion point we pick $\delta \in [q] \setminus \{0\}$ and make the inserted symbol equal to $\delta$ plus its successor.

The size of the parameter set for the construction under the first restriction is $q^l\binom{l-1}{s}\binom{s}{a}(q-1)^s$, which it very similar to the asymptotic upper bound of Lemma~\ref{lemma:edge-ub}.

\begin{definition}
A string is alternating if some $u \in [q]$ appears at all even indices, some $v \in [q]$ appears at all odd indices, and $u \neq v$.
Let $A_{q,n}$ be the set of nonalternating $q$-ary strings of length $n$.
\end{definition}
The empty string and all strings of length one are trivially alternating, so the shortest nonalternating strings have length two.
For each length $n \geq 2$, each of the $q$ choices for $u$ and $q-1$ choices for $v$ results in a unique string, so $|A_{q,n}| =q^n - q(q-1)$.

To explain the purpose of the second restriction, we must first describe the deconstruction procedure.
Start with an edge $(x,y)$.
Beginning at the left, find the longest matching prefix of $x$ and $y$ and delete it from both.
This prefix is the first interval of $z$.
Now the first symbols of $x$ and $y$ differ.
One of these symbols is part of the next interval of $z$ and the other was an insertion, but we do not know which is which.

To resolve this situation, apply the following heuristic.
Delete the first symbol of $x$ and determine the length of the longest common prefix of $y$ and the rest of $x$.
Then do the same with the roles of $x$ and $y$ reversed.
Assume that the deleted symbol that resulted in the longer common prefix was the insertion and that the longer prefix was the next interval of $z$.
After removing this prefix, either the first symbols of $x$ and $y$ again differ or $x$ and $y$ are both the empty string.
Apply this heuristic until the latter case is achieved. 

We will show that this heuristic is always correct when applied to edges produced under the second restriction.

\subsection{Formalization of the construction and deconstruction functions}
Our construction function, \textsc{Construct}, is specified in Algorithm~\ref{alg:con} and our deconstruction function, \textsc{Deconstruct}, is specified in Algorithm~\ref{alg:decon}.
Example of the construction and deconstruction algorithms are provided in Figures~\ref{figure:con} and \ref{figure:decon}.

\begin{figure}
\begin{IEEEeqnarray*}{l}
\textsc{Construct}(11,(L,1,102),(R,2,21211),(L,2,021))\\
\quad 
\begin{IEEEeqnarraybox}{rlCl}
\textsc{Insert}&(L,1,102) &=& \rstack{2102}{102}\\
\textsc{Insert}&(R,2,21211) &=& \rstack{21211}{121211}\\
\textsc{Insert}&(L,2,021) &=& \rstack{2021}{021}\\
\end{IEEEeqnarraybox}\\
= \rstack{11}{11} \rstack{2102}{102} \rstack{21211}{121211} \rstack{2021}{021}=\stack{112102212112021}{11102121211021}
\end{IEEEeqnarray*}
\caption{
An example of the construction procedure for a pair of strings.
The \textsc{Insert} function is applied to each triple $(\text{LR} \times ([q] \setminus \{0\}) \times [q]^*)$ to produce a pair of string segments.
\textsc{Construct} concatenates these to produce the final pair.
}
\label{figure:con}
\end{figure}

\begin{figure}
\begin{IEEEeqnarray*}{l}
\textsc{Deconstruct}\wrap{\stack{112102212112021}{11102121211021}}\\
\quad 
\begin{IEEEeqnarraybox}{l}
\textsc{Match}\wrap{\stack{112102212112021}{11102121211021}} = 11 \stack{2102212112021}{102121211021}\\
\textsc{Delete}\wrap{\stack{2102212112021}{102121211021}}\\
\quad 
\begin{IEEEeqnarraybox}{rlCrl}
\textsc{Match}&\wrap{\stack{102212112021}{102121211021}} &=& 102 & \stack{212112021}{121211021} \checkmark\\
\textsc{Match}&\wrap{\stack{2102212112021}{02121211021}} &=& \epsilon & \stack{2102212112021}{02121211021}\\
\end{IEEEeqnarraybox}\\
=(L,1,102) \stack{212112021}{121211021}\\
\textsc{Delete}\wrap{\stack{212112021}{121211021}}\\
\quad 
\begin{IEEEeqnarraybox}{rlCrl}
\textsc{Match}&\wrap{\stack{12112021}{121211021}} &=& 121 & \stack{12021}{211021}\\
\textsc{Match}&\wrap{\stack{212112021}{21211021}} &=& 21211 & \stack{2021}{021} \checkmark\\
\end{IEEEeqnarraybox}\\
= (R,2,21211) \stack{2021}{021}\\
\textsc{Delete}\wrap{\stack{2021}{021}}\\
\quad  
\begin{IEEEeqnarraybox}{rlCrl}
\textsc{Match}&\wrap{\stack{021}{021}} &=& 021 & \stack{$\epsilon$}{$\epsilon$}\checkmark\\
\textsc{Match}&\wrap{\stack{2021}{21}} &=& 2 & \stack{021}{1}\\
\end{IEEEeqnarraybox}\\
= (L,2,021) \stack{$\epsilon$}{$\epsilon$}\\
\end{IEEEeqnarraybox}\\
= 11,(L,1,102),(R,2,21211),(L,2,021)\\
\end{IEEEeqnarray*}
\caption{
An example of the deconstruction process.
First, \textsc{Match} strips off the common prefix.
The \textsc{Delete} function tests whether it a longer common prefix is achieved by deleting the head of the first string or the second string.
The check marks indicate the longer match.
It produces a triple specifying that deletion and prefix.
}
\label{figure:decon}
\end{figure}

The functions treat strings as lists of symbols.
We represent the empty list as $\epsilon$.
We write the concatenation of $x$ and $y$ as $x \con y$.
The function \textsc{Head} returns the first symbol of a nonempty list and the function \textsc{Tail} returns everything except the head.
The function \textsc{Length} returns the number of symbols in the string.

The \textsc{Construct} function produces a pair of strings.
As its input, \textsc{Construct} takes $s+1$ intervals of arbitrary lengths, a subset of $[s]$, and $s$ nonzero $q$-ary symbols.
Let $\textsc{LR} = \{\textsc{Left}, \textsc{Right}\}$.
We represent the subset $T \subseteq [s]$ as a string $t \in \textsc{LR}^s$, where $t_i = \textsc{Left}$ if $i \in T$ and $t_i = \textsc{Right}$ if $i \not\in T$.
Thus the input to \textsc{Construct} is an element of 
\begin{equation*}
([q]^*)^{s+1} \times \text{LR}^{s} \times ([q] \setminus \{0\})^s = [q]^* \times \left(\text{LR} \times ([q] \setminus \{0\}) \times [q]^*\right)^s.
\end{equation*}

The \textsc{Insert} function takes one of the triples $\left(\text{LR} \times ([q] \setminus \{0\}) \times [q]^*\right)$ as an argument and outputs two strings.
Let $w$ be the string from the triple.
One of the output strings is $w$ and the other is $w$ with a single symbol has been inserted at the head.
\textsc{Construct} applies \textsc{Insert} to each triple, concatenates the results, and prepends the remaining input string to each output.

The \textsc{Match} function takes two strings $x$ and $y$, finds their longest common prefix, and outputs the prefix and the two corresponding suffixes.
The \textsc{Deconstruct} uses \textsc{Match} to remove the common prefix of the input strings, then repeatedly calls \textsc{Delete}.
\textsc{Delete} takes a pair of strings $x$ and $y$ that differ in their first symbol and each application of \textsc{Delete} undoes the effect of an \textsc{Insert}.
\textsc{Delete} calls \textsc{Match} on $(\textsc{Tail}(x), y)$ and on $(x,\textsc{Tail}(y))$ and then preforms the deletion that resulted in a longer common prefix.
The information about the deletion and prefix become a triple $\left(\text{LR} \times ([q] \setminus \{0\}) \times [q]^*\right)$.
\textsc{Delete} returns this triple along with two suffixes from the match.

\algblockdefx[TypedFunction]{Typed}{EndTyped}%
[3]{\textsc{#1} : #2 \\ \textsc{#1}(#3)}%
{}
\newfloat{algorithm}{t}{h}
\begin{algorithm}
\caption{Construct an edge}
\begin{algorithmic}
\Typed{Construct}{$[q]^* \times (\text{LR} \times ([q] \setminus \{0\}) \times [q]^*)^s \rightarrow [q]^* \times [q]^*$}{$w_0,t$}
\State $(x,y) \gets \textsc{C}(t)$
\State \Return $(w_0:x,w_0:y)$
\EndTyped
\end{algorithmic}

\begin{algorithmic}
\Typed{C}{$(\text{LR} \times ([q] \setminus \{0\}) \times [q]^*)^s \rightarrow [q]^* \times [q]^*$}{$t$}
\If{$t = \epsilon$}
 \State \Return $(\epsilon,\epsilon)$
\Else
 \State $(u,v) \gets \textsc{Insert}(\textsc{Head}(t))$
 \State $(x,y) \gets \textsc{C}(\textsc{Tail}(t))$
 \State \Return $(u : x, v : y)$
\EndIf
\EndTyped
\end{algorithmic}

\begin{algorithmic}
\Typed{Insert}{$\text{LR} \times ([q] \setminus \{0\}) \times [q]^* \rightarrow [q]^* \times [q]^*$}{$lr,\delta,w$}
\State $w' \gets (\delta + \textsc{Head}(w)):w$
\If{$lr = \textsc{Left}$}
 \State \Return $(w', w)$
\Else
 \State \Return $(w, w')$
\EndIf
\EndTyped
\end{algorithmic}
\label{alg:con}
\end{algorithm}

\begin{algorithm}
\caption{Deconstruct an edge}
\begin{algorithmic}
\Typed{Deconstruct}{$[q]^* \times [q]^* \rightarrow [q]^* \times (\text{LR} \times ([q] \setminus \{0\}) \times [q]^*)^s$}{$x,y$}
\State $(w_0,x,y) \gets \textsc{Match}(x,y)$
\State \Return $(w_0,\textsc{D}(x,y))$
\EndTyped
\end{algorithmic}

\begin{algorithmic}
\Typed{D}{$[q]^* \times [q]^* \rightarrow (\text{LR} \times ([q] \setminus \{0\}) \times [q]^*)^s$}{$x,y$}
\If{$x = \epsilon \vee y = \epsilon$}
 \State \textbf{assert} $x = \epsilon \wedge y = \epsilon$
 \State \Return $\epsilon$
\Else
 \State $(w,x,y) \gets \textsc{Delete}(x,y)$
 \State \Return $(w:\textsc{D}(x,y))$
\EndIf
\EndTyped
\end{algorithmic}

\begin{algorithmic}
\Typed{Delete}{$[q]^* \times [q]^* \rightarrow (\text{LR} \times ([q] \setminus \{0\}) \times [q]^*) \times [q]^* \times [q]^*$}{$x,y$}
\State $g = \textsc{Head}(x) - \textsc{Head}(y)$
\State $(a,b,c) \gets \textsc{Match}(\textsc{Tail}(x),y)$
\State $(d,e,f) \gets \textsc{Match}(x,\textsc{Tail}(y))$
\State \textbf{assert} $\textsc{Length}(a) \neq \textsc{Length}(d)$
\If{$\textsc{Length}(a) > \textsc{Length}(d)$}
 \State \Return $((\textsc{Left},g,a),b,c)$
\Else
 \State \Return $((\textsc{Right},(-g),d),e,f)$
\EndIf

\EndTyped
\end{algorithmic}

\begin{algorithmic}
\Typed{Match}{$[q]^i \times [q]^j \rightarrow [q]^k \times [q]^{i-k} \times [q]^{j-k}$}{$x,y$}
\State $w \gets \epsilon$
\While{$x \neq \epsilon \wedge y \neq \epsilon \wedge \textsc{Head}(x) = \textsc{Head}(y)$}
 \State $w \gets w : \textsc{Head}(x)$
 \State $x \gets \textsc{Tail}(x)$
 \State $y \gets \textsc{Tail}(y)$
\EndWhile
\State \Return $(w, x, y)$
\EndTyped
\end{algorithmic}

\label{alg:decon}
\end{algorithm}

\subsection{Deconstruction}
Now we will show that \textsc{Deconstruct} is a left inverse of \textsc{Construct}.
The first step is to look at the inner functions: \textsc{Insert} and \textsc{Delete}.
\begin{lemma}
\label{lemma:insert-delete}
For $lr \in \text{LR}$, $\delta \in [q] \setminus \{0\}$, and $w \in A_{q,m}$, let $(x,y) = \textsc{Insert}(lr,\delta,w)$.
Let $u, v \in [q]^*$ such that if both are nonempty, they have different first symbols.
Then $\textsc{Delete}(x \con u,y \con v) = ((lr,\delta,w),u,v)$.
\end{lemma}
\begin{IEEEproof}
Let $w = w_0^{m-1} = (w_0,w_1,\dots,w_{m-1})$.
Without loss of generality let $lr = \textsc{Left}$, so $x=(w_0+\delta) \con w$ and $y=w$.
First, $\textsc{Delete}(x \con u,y \con v)$ computes $g = (w_0+\delta)-w_0 = \delta$.
Next, it evaluates $\textsc{Match}(w \con u,w \con v) = (w,u,v)$ because either $u_0 \neq v_0$ or one of $u$ and $v$ is the empty string.
Thus the length of the first match is $\textsc{Length}(w)=m$.
Second, it evaluates $\textsc{Match}((w_0+\delta) \con w \con u, w_1^{m-1} \con v)$.
If the length of this matches is at least $m-1$, then $w_0 + \delta = w_1$ and $w_i=w_{i+2}$ for $0 \leq i \leq m-3$.
This would make $w$ alternating, so the lengths of the second match is at most $m-2$.
The first match is longer than the second, so the first branch of the if statement is taken and the function returns $((\textsc{Left}, \delta, w), u,v)$.
\end{IEEEproof}

\begin{definition}
For all $q,l,a,b \in \mathbb{N}$, let $s=a+b$.
Let $P_{q,l,s}$ be the set
\begin{equation*}
\bigcup_{\mathbf{c} \in M(s+1,l,2)} A_{q,c_0} \times \prod_{i=1}^{s} \left(\textsc{LR} \times ([q] \setminus \{0\}) \times A_{q,c_i}\right)
\end{equation*}
and let $P_{q,l,a,b}$ be the subset of $P_{q,l,s}$ with exactly $a$ appearances of \textsc{Left}.
\end{definition}

\begin{lemma}
\label{lemma:inverse-functions}
For all $q,l,s \in \mathbb{N}$ and $p \in P_{q,l,s}$, $\textsc{Deconstruct}(\textsc{Construct}(p)) = p$.
\end{lemma}
\begin{IEEEproof}
Let $p = (w_0,t_s,\dots,t_1)$ where $t_i = (lr_i,\delta_i,w_i)$.
The initial call to \textsc{Match} in \textsc{Deconstruct} finds $w_0$, so $\textsc{Deconstruct}(\textsc{Construct}(p)) = (w_0,\textsc{D}(\textsc{C}(t_1,\dots,t_s)))$.
We show that $\textsc{D}(\textsc{C}(t_s,\dots,t_1))) = (t_1,\dots,t_s)$ by induction.
For the base case, $\textsc{D}(\textsc{C}(\epsilon)) = \textsc{D}(\epsilon, \epsilon) = \epsilon$.
For the induction step, note that $(u,v) = \textsc{C}(t_i,\dots,t_1)$ can be taken to be the $u$ and $v$ in the statement of Lemma~\ref{lemma:insert-delete} because they are either both empty of they have different first symbols.
Then Lemma~\ref{lemma:insert-delete} gives $\textsc{D}(\textsc{C}(t_{i+1},\dots,t_0)) = t_{i+1} \con \textsc{D}(\textsc{C}(t_i,\dots,t_0))$.
\end{IEEEproof}

\begin{lemma}
\label{lemma:construct-edge}
For all $q,l,a,b \in \mathbb{N}$, $s=a+b$, and $p \in P_{q,l,a,b}$, $\textsc{Construct}(p) \in E(B_{q,l,a,b})$.
\end{lemma}
\begin{IEEEproof}
Let $(x,y) = \textsc{Construct}(p)$.
Let $p = (w_0,t_1,\dots,t_s)$ where $t_i = (lr_i,\delta_i,w_i)$.
One output of $\textsc{Insert}(lr_i,\delta_i,w_i)$ is a strict superstring $w_i$ and the other is $w_i$.
Thus both $x$ and $y$ are superstrings of $w_0 \con w_1 \con \dots \con w_s$.
The longer output of \textsc{Insert} becomes part of $x$ $a$ times, so the length of $x$ is $l+a$.
Similarly the length of $y$ is $l+b$. 
\end{IEEEproof}
\subsection{The lower bound}

\begin{lemma}
\label{lemma:parameter-asymptotics}
For fixed $q,a,b \in \mathbb{N}$, 
$|P_{q,l,a,b}| \gtrsim q^l \binom{l}{s}\binom{s}{a}(q-1)^s$.
\end{lemma}
\begin{IEEEproof}
Refactor $P_{q,l,s}$ as 
\begin{equation*}
\textsc{LR}^s \times ([q] \setminus \{0\})^s \times \bigcup_{\lambda \in M(s+1,l,2)} \prod_{i=0}^{s} A_{q,c_i}.
\end{equation*}
In $P_{q,l,a,b}$, the element of $\textsc{LR}^s$ is one of the $\binom{s}{a}$ strings with exactly $a$ appearances of $\textsc{Left}$.
There are $(q-1)^s$ possibilities for $([q] \setminus \{0\})^s$.
For $\lambda_i \geq 2$, $|A_{q,\lambda_i}| = q^{\lambda_i} - q(q-1)$, so the size the union is
\begin{IEEEeqnarray*}{Cl}
&\sum_{\lambda \in M(s+1,l,2)} \prod_{i=0}^s (q^{\lambda_i}-q(q-1))\\
\geq&\sum_{\lambda \in M(s+1,l,2)} \prod_{i=0}^s (q^{\lambda_i}-q^2)\\
=&q^l\sum_{\lambda \in M(s+1,l,2)} \prod_{i=0}^s \left(1-q^{2-\lambda_i}\right)\\
\geq&q^l\sum_{\lambda \in M(s+1,l,2+\log_q l)} \prod_{i=0}^s \left(1-q^{2-\lambda_i}\right)\\
\geq&q^l\binom{l-(1+\log_q l)(s+1)-1}{s} (1-l^{-1})^{s+1}\\
\sim&q^l \binom{l}{s}.
\end{IEEEeqnarray*}
Thus $|P_{q,l,a,b}| \gtrsim q^l \binom{l}{s}\binom{s}{a}(q-1)^s$.
\end{IEEEproof}
Our bounds establish the asymptotic growth of the number of edges.
\begin{theorem}
\label{theorem:edges}
For fixed $q,a,b \in \mathbb{N}$, the number of edges in $B_{q,l,a,b}$ satisfies $|E(B_{q,l,a,b})| \sim q^l \binom{l}{s}\binom{s}{a}(q-1)^s$.
The average of $S_{a,b}(x)$ over all $x \in [q]^n$ is asymptotic to $\binom{n}{s}\binom{s}{a}(q-1)^{s}q^{-a}$.
\end{theorem}
\begin{IEEEproof}
From Lemma~\ref{lemma:inverse-functions} and Lemma~\ref{lemma:construct-edge}, $|E(B_{q,l,a,b})| \geq |P_{q,l,a,b}|$.
Lemma~\ref{lemma:edge-ub} provides the asymptotic upper bound and Lemma~\ref{lemma:parameter-asymptotics} provides the asymptotic lower bound.

For $x \in [q]^n$, the set $S_{a,b}(x)$ is the neighborhood of $x$ in $B_{q,n-a,a,b}$.
Each edge involves exactly one of the $q^n$ left vertices and $\binom{n-a}{a} \sim \binom{n}{a}$.
\end{IEEEproof}
Now we can conclude that most edges are constructable by our method.
This is a necessary condition for the asymptotic tightness of our ultimate lower bound on input degree.

\section{Bounds on Input Degree and Code Size}
\label{section:bounds}
\begin{lemma}
\label{lemma:degree-lb}
Let $x \in [q]^n$ be a string with $r$ runs. Let $c$ be the length of the longest alternating interval of $x$. Then $|S_{a,b}(x)|$, the number of unique strings that can be produced from $x$ by $a$ deletions and $b$ insertions, is at least
\begin{equation*}
\binom{r - (a+1)(c+1)}{a}\binom{n - 1 - 2a(c+1) - (b+1)c}{b}(q-1)^b.
\end{equation*}
\end{lemma}
\begin{IEEEproof}
For each $x \in [q]^n$, we identify a subset $P_x \subseteq P_{q,n-a,a,b}$ such that for all $p \in P_x$, $\textsc{Construct}(p) = (x,y)$.
From Lemma~\ref{lemma:construct-edge}, all $y$ produced this way are in $S_{a,b}(x)$.
From Lemma~\ref{lemma:inverse-functions}, $|S_{a,b}(x)| \geq |P_x|$.

To produce an element of $P_x$, we select $a$ symbols of $x$ for deletion, select $b$ spaces in $x$ for insertion, and specify the $b$ new symbols.
The symbols selected of deletion and the spaces selected for insertion partition $x$ into $s+1$ intervals.
To ensure that none of these intervals are alternating, we will require that all of the intervals contain at least $c+1$ symbols.

There are many equivalent ways to extends a run by inserting a matching symbol.
\textsc{Construct} extends runs by adding a symbol at the right end, so we only select symbols for deletion from those at the right end of a run.
We need there to be at least $c+1$ symbols between consecutive deleted symbols.
It is easier to enforce the stronger condition that there are $c+1$ end of run symbols between consecutive deleted symbols.
There are $\binom{r - (a+1)(c+1)}{a}$ ways to pick the symbols for deletion that satisfy this condition.

There are $n-1$ potential spaces in which an insertion can be made.
Insertions cannot be performed in the $c+1$ spaces before and after a deleted symbol.
In the worst case, all of these forbidden spaces are distinct, leaving $n-1-2a(c+1)$ spaces to choose from.
There must be $c+1$ symbols between any two consecutive chosen spaces, before the first chosen space, and after the last chosen space.
Thus there must be at least $c$ spaces in each of these $b+1$ intervals.
Again, it is easier to enforce the stronger condition that there are at least $c$ spaces not near a deletion in each interval.
Thus there are always at least $\binom{n - 1 - 2a(c+1) - (b+1)c}{b}$ ways to pick the spaces.

Finally, for each of the $b$ insertion points, we must specify the difference inserted symbol and its successor.
Thus, there are $(q-1)^b$ choices for this step.
\end{IEEEproof}
The following argument, very similar to Lemma~\ref{lemma:edge-ub}, shows that this degree lower bound is asymptotically tight.
This is a generalization of a lemma of Levenshtein~\cite{levenshtein_binary_1966}, 
\begin{lemma}
\label{lemma:degree-ub}
For all $q,n,r,a,b \in \mathbb{N}$ with $s=a+b$, if $x \in [q]^n$ has $r$ runs, then
\begin{equation*}
|S_{a,b}(x)| \leq \binom{r+a-1}{a}I_{q,b,n-a+b}.
\end{equation*}
\end{lemma}
\begin{IEEEproof}
Any substring of $x$ can be the number of symbols deleted from each run. 
This is a composition of $a$ with $r$ parts, so $|S_{a,0}(x)| \leq |M(a,r,0)| = \binom{r-1+a}{r-1} = \binom{r+a-1}{a}$.
Each string in $S_{a,b}(x)$ is a superstring of one of these substrings.
Each substring has exactly $I_{q,b,n-a+b}$ superstrings of length $n-a+b$.
\end{IEEEproof}
If $r=pn$ for fixed $p$, both bounds are asymptotic to 
\begin{equation*}
\binom{r}{a}\binom{n}{b}(q-1)^b.
\end{equation*}

To apply Lemma~\ref{lemma:degree-lb} to a string, we need two statistics of that string: the number of runs and the length of the longest alternating interval.
The next two lemmas concern the distributions of these statistics. 
\begin{lemma}
\label{lemma:alternating-ub}
The number of $q$-ary strings of length $n$ with an alternating interval of length at least $c$ is at most $(n-c+1)q^{n-c+1}(q-1)$ .
\end{lemma}
\begin{IEEEproof}
A string of length $n$ contains $n-c+1$ intervals of length $c$.
If some interval of length at least $c$ is alternating, at least one of intervals of length exactly $c$ is alternating.
There are $q(q-1)$ choices for the symbols in the alternating interval and $q^{n-c}$ choices for the remaining symbols.
\end{IEEEproof}

\begin{lemma}
\label{lemma:runs-ub}
The number of $q$-ary strings of length $n$ with $\left(\frac{q-1}{q}-\epsilon\right)(n-1) + 1$ or fewer runs is at most $q^ne^{-2(n-1)\epsilon^2}$.
\end{lemma}
\begin{IEEEproof}
For $x \in [q]^n$, let $x' \in [q]^{n-1}$ be the string of first differences of $x$.
That is, let $x'_i = x_{i+1} - x_i \bmod q$.
If $x$ has $r$ runs, then $x'_i$ is nonzero at the $r-1$ boundaries between runs.
Thus there are $q\binom{n-1}{r-1}(q-1)^{r-1}$ strings with exactly $r$ runs.
The number of strings with few runs is 
\begin{IEEEeqnarray*}{Cl}
&\sum_{i=0}^{\left(\frac{q-1}{q}-\epsilon\right)(n-1)} \binom{n-1}{i}(q-1)^i\\
=& q^{n-1} \sum_{i=0}^{\left(\frac{q-1}{q}-\epsilon\right)(n-1)} \binom{n-1}{i}\left(\frac{q-1}{q}\right)^i\left(\frac{1}{q}\right)^{n-1-i}\\  
\leq& q^{n-1}e^{-2(n-1)\epsilon^2}.
\end{IEEEeqnarray*}
The upper bound comes from the application of Hoeffding's inequality to the binomial distribution~\cite{hoeffding_probability_1963}.
\end{IEEEproof}

Now we have all of the ingredients required to execute the strategy described in Section~\ref{section:motivation}.
\begin{lemma}
\label{lemma:classes}
Let $q,a,b \in \mathbb{N}$ be fixed and let $s = a+b$. For all $t \in \mathbb{N}$, there is a sequence of subsets $T_n \subseteq [q]^n$ such that $|T_n|$ is $O(q^n/n^t)$ and 
\begin{equation*}
\min_{x \in [q]^n \setminus T_n} |S_{a,b}(x)| \gtrsim \frac{(q-1)^s}{q^a}\binom{n}{s}\binom{s}{b}
\end{equation*}
\end{lemma}
\begin{IEEEproof}
We form two classes of bad strings: strings with a long alternating interval, and strings with few runs. 
Call these classes $T_n'$ and $T_n''$ respectively.
Let $T_n = T_n' \cup T_n''$.

A string falls into $T_n'$ if it has an alternating subinterval of length at least $c$. 
If we let $c = (t+1) \log_q n$, then by Lemma~\ref{lemma:alternating-ub} we have 
\begin{equation*}
|T_n'| <nq^{n-c+1}(q-1)=n^{-t}q^{n+1}(q-1)
\end{equation*}
which is $O(q^n/n^t)$.
Over all strings in $[q]^n$, the average number of runs is $\frac{q-1}{q}~(n~-~1)~+~1$.
A string falls into $T_n''$ if it has at most $\left(\frac{q-1}{q} - \epsilon \right)(n-1)+1$ runs.
If we let $\epsilon = \sqrt{\frac{t\log n}{2(n-1)}}$, then by Lemma~\ref{lemma:runs-ub} we have  
\begin{equation*}
|T_n''| \leq q^ne^{-2(n-1)\epsilon^2} = q^ne^{-t \log n} = q^n/n^t.
\end{equation*}
For fixed $t$, this $\epsilon$ is $o(1)$, so
$\left(\frac{q-1}{q} - \epsilon \right)(n-1)+1 \sim \frac{(q-1)n}{q}.$

Now we can apply Lemma~\ref{lemma:degree-lb} to lower bound the degree of the strings in $[q]^n \setminus T_n$.
The first multiplicative term in the lower bound is asymptotic to
\begin{IEEEeqnarray*}{rCl}
\binom{\frac{q-1}{q}n -(a+1)((t+1)\log_q n +1)}{a}
&\sim& \binom{\frac{q-1}{q}n}{a}\\
&\sim& \left(\frac{q-1}{q}\right)^a\binom{n}{a}.
\end{IEEEeqnarray*}
The second term is asymptotic to
\begin{equation*}
\binom{n-1-2a-(2a+b+1)(t+1)\log_q n}{b} \sim \binom{n}{b}.
\end{equation*}
Thus
\begin{IEEEeqnarray*}{rCl}
\min_{x \in [q]^n \setminus T_n} |S_{a,b}(x)| &\gtrsim&\left(\frac{q-1}{q}\right)^a\binom{n}{a}\binom{n}{b}(q-1)^b \\
&\sim& \frac{(q-1)^s}{q^a}\binom{n}{s}\binom{s}{b}.
\end{IEEEeqnarray*}
\end{IEEEproof}

Our main theorem follows easily.
\begin{theorem}
\label{theorem:main}
For fixed $q,s \in \mathbb{N}$, the number of codewords in an $n$-symbol $q$-ary $s$-deletion correcting code satisfies
\begin{equation*}
C_{q,s,n} \lesssim \min_{0 \leq b \leq s} \frac{q^{n+b}}{(q-1)^s\binom{n}{s}\binom{s}{b}}.
\end{equation*}%
\end{theorem}
\begin{IEEEproof}
Consider an $a$-deletion $b$-insertion channel with $a+b=s$.
By Lemma~\ref{lemma:channel-equiv}, any code for this channel can also correct $s$ deletions.
There are $q^{n-a+b}$ possible outputs, so for any $T_n~\subseteq~[q]^n$,
\begin{equation*}
C_{q,s,n} \lesssim \frac{q^{n-a+b}}{\min_{x \in [q]^n \setminus T_n} |S_{a,b}(x)|} + |T_n|.
\end{equation*}
By setting $t = s+1$ in Lemma~\ref{lemma:classes} we obtain an asymptotic upper bound of
\begin{equation*}
C_{q,s,n} \lesssim \frac{q^{n-a+b}}{\frac{(q-1)^s}{q^a}\binom{n}{s}\binom{s}{b}} + O \left(\frac{q^n}{n^{s+1}}\right) \sim \frac{q^{n+b}}{(q-1)^s\binom{n}{s}\binom{s}{b}}.
\end{equation*}
\end{IEEEproof}
This improves \eqref{eq:lev-upper}, Levenshtein's upper bound, by a factor of $\binom{s}{b}q^{-b}$.
By setting $b$ to zero we recover Levenshtein's bound.
Whenever $s>q$, $\binom{s}{1}q^{-1}>\binom{s}{0}q^0 = 1$ so setting $b$ to one in the generalized bound offers an improvement.

\begin{corollary}
\label{corollary:opt}
If $q+1$ divides $s$, the size of a $q$-ary $s$-deletion correcting code satisfies
\begin{equation*}
C_{q,s,n} \lesssim \frac{3\sqrt{s}q^{n+s+\frac{1}{2}}}{(q+1)^{s+1}(q-1)^s\binom{n}{s}}.
\end{equation*}
\end{corollary}
\begin{IEEEproof}
We optimize over $b$ in Theorem~\ref{theorem:main}.
The factor $\binom{s}{b}q^{-b}$ is a constant times a binomial distribution:
\begin{equation*}
\label{eq:improvement}
\left(\frac{q+1}{q}\right)^s\binom{s}{b}\left(\frac{1}{q+1}\right)^b\left(\frac{q}{q+1}\right)^{s-b}.
\end{equation*}
The maximum is achieved by $b = \left\lfloor \frac{s+1}{q+1} \right\rfloor$.
When $q+1$ divides $s$, the maximum is at least 
\begin{equation*}
\left(\frac{q+1}{q}\right)^s\frac{1}{3}\sqrt{\frac{q+1}{q s/(q+1)}} = \frac{(q+1)^{s+1}}{3 \sqrt{s} q^{s+\frac{1}{2}}}
\end{equation*}
by Stirling's approximation.
See Appendix~\ref{section:stirling} for details.
\end{IEEEproof}

\section{Concluding remarks}
In this paper, we extended Levenshtein's strategy for obtaining an upper bound on the size of deletion codes.
Levenshtein's bound arises from the deletion channel.
We derived the corresponding bounds from channels that perform a mixture of deletions and insertions.
This results in an improvement whenever the number of errors, $s$, is larger than the alphabet size, $q$.
The best version of our bound uses a channel where the ratio of deletions to insertions is $q$ to one.

Our argument relies on the fact that the channel graphs are approximately regular in the asymptotic regime where the number of errors is fixed.
A natural question is whether it can be extended to the regime where the number of errors is a constant fraction of the input length.
However, it is not clear whether the graphs are approximately regular in the latter regime.
The argument of this paper relies on the typical distance between errors going to infinity.
Any interaction between two errors, which occurs via an alternating interval, becomes rare.
When the typical distance does not grow with input length, interactions will not be rare and it will not be possible to simply discard the cases where they occur.
Instead it will be necessary to understand the details of more types of interactions between errors. 

\appendices
\section{}
\label{section:stirling}
\blockcomment{
\begin{lemma}
\label{lemma:stirling}
For $a,b,n \in \mathbb{N}$,
\begin{equation*}
\binom{(a+b)n}{an} \left(\frac{a}{a+b}\right)^{an}\left(\frac{b}{a+b}\right)^{bn} \geq \frac{1}{3}\sqrt{\frac{a+b}{a b n}}.
\end{equation*}
\end{lemma}
\begin{IEEEproof}}
One form of Stirling's approximation is \cite{feller_introduction_2008} 
\begin{equation*}
\sqrt{2\pi} \leq \frac{n!}{\sqrt{n}\left(\frac{n}{e}\right)^n} \leq e.
\end{equation*}
Then for $\alpha, \beta, n \in \mathbb{N}$, consider the binomial distribution produced by $(\alpha + \beta)n$ trials and success probability $\alpha/(\alpha + \beta)$.
The most likely outcome is $\alpha n$ successes and the probability of that outcome is:
\begin{IEEEeqnarray*}{Cl}
& \max_i \binom{(\alpha + \beta)n}{i} \left(\frac{\alpha}{\alpha + \beta}\right)^i \left(\frac{\beta}{\alpha + \beta}\right)^{(\alpha +\beta) n - i} \\
=&\binom{(\alpha + \beta)n}{\alpha n} \left(\frac{\alpha}{\alpha + \beta}\right)^{\alpha n} \left(\frac{\beta}{\alpha + \beta}\right)^{\beta n} \\
\geq & \frac{\sqrt{2\pi (\alpha + \beta)n}\left(\frac{(\alpha + \beta)n}{e}\right)^{(\alpha + \beta)n}}{e \sqrt{\alpha n}\left(\frac{\alpha n}{e}\right)^{\alpha n} e \sqrt{\beta n}\left(\frac{\beta n}{e}\right)^{\beta n}}
\frac{\alpha^{\alpha n}\beta^{\beta n}}{(\alpha + \beta)^{(\alpha + \beta)n}} \\
= & \frac{\sqrt{2 \pi}}{e^2}\sqrt{\frac{\alpha + \beta}{\alpha \beta n}} \\
\geq & \frac{1}{3}\sqrt{\frac{\alpha + \beta}{\alpha \beta n}}.
\end{IEEEeqnarray*}
\bibliographystyle{IEEEtran} 
\bibliography{IEEEabrv,cullina}

\begin{thebibliography}{10}
\providecommand{\url}[1]{#1}
\csname url@rmstyle\endcsname
\providecommand{\newblock}{\relax}
\providecommand{\bibinfo}[2]{#2}
\providecommand\BIBentrySTDinterwordspacing{\spaceskip=0pt\relax}
\providecommand\BIBentryALTinterwordstretchfactor{4}
\providecommand\BIBentryALTinterwordspacing{\spaceskip=\fontdimen2\font plus
\BIBentryALTinterwordstretchfactor\fontdimen3\font minus
  \fontdimen4\font\relax}
\providecommand\BIBforeignlanguage[2]{{%
\expandafter\ifx\csname l@#1\endcsname\relax
\typeout{** WARNING: IEEEtran.bst: No hyphenation pattern has been}%
\typeout{** loaded for the language `#1'. Using the pattern for}%
\typeout{** the default language instead.}%
\else
\language=\csname l@#1\endcsname
\fi
#2}}

\bibitem{calabi_general_1969}
L.~Calabi and W.~E. Hartnett, ``Some general results of coding theory with
  applications to the study of codes for the correction of synchronization
  errors*,'' \emph{Information and Control}, vol.~15, no.~3, p. 235–249,
  1969.

\bibitem{cullina_improvement_2013}
D.~Cullina and N.~Kiyavash, ``An improvement to levenshtein's upper bound on
  the cardinality of deletion correcting codes,'' in \emph{{IEEE} International
  Symposium on Information Theory Proceedings}, July 2013.

\bibitem{cullina_coloring_2012}
D.~Cullina, A.~Kulkarni, and N.~Kiyavash, ``A coloring approach to constructing
  deletion correcting codes from constant weight subgraphs,'' in \emph{{IEEE}
  International Symposium on Information Theory Proceedings ({ISIT)}}, July
  2012, p. 513 –517.

\bibitem{feller_introduction_2008}
W.~Feller, \emph{An introduction to probability theory and its
  applications}.\hskip 1em plus 0.5em minus 0.4em\relax John Wiley \& Sons,
  2008, vol.~2.

\bibitem{hirschberg_bounds_1999}
D.~Hirschberg, ``Bounds on the number of string subsequences,'' in
  \emph{Combinatorial Pattern Matching}, 1999, p. 115–122.

\bibitem{hoeffding_probability_1963}
\BIBentryALTinterwordspacing
W.~Hoeffding, ``Probability inequalities for sums of bounded random
  variables,'' \emph{Journal of the American Statistical Association}, vol.~58,
  no. 301, pp. 13--30, Mar. 1963, {ArticleType:} research-article / Full
  publication date: Mar., 1963 / Copyright © 1963 American Statistical
  Association. [Online]. Available: \url{http://www.jstor.org/stable/2282952}
\BIBentrySTDinterwordspacing

\bibitem{kulkarni_non-asymptotic_2012}
\BIBentryALTinterwordspacing
A.~A. Kulkarni and N.~Kiyavash, ``Non-asymptotic upper bounds for deletion
  correcting codes,'' \emph{{IEEE} Transactions on Information Theory}, 2012.
  [Online]. Available: \url{http://arxiv.org/abs/1211.3128}
\BIBentrySTDinterwordspacing

\bibitem{levenshtein_binary_1966}
V.~I. Levenshtein, ``Binary codes capable of correcting deletions, insertions,
  and reversals,'' in \emph{Soviet physics doklady}, vol.~10, 1966, p.
  707–710.

\bibitem{levenshtein_elements_1974}
------, ``Elements of coding theory,'' \emph{Diskretnaya matematika i
  matematicheskie voprosy kibernetiki}, p. 207–305, 1974.

\bibitem{levenshtein_bounds_2002}
------, ``Bounds for deletion/insertion correcting codes,'' in \emph{{IEEE}
  International Symposium on Information Theory Proceedings}, 2002, p. 370.

\bibitem{liron_characterization_2012}
Y.~Liron and M.~Langberg, ``A characterization of the number of subsequences
  obtained via the deletion channel,'' in \emph{{IEEE} International Symposium
  on Information Theory Proceedings}, 2012, p. 503–507.

\bibitem{mercier_number_2008}
H.~Mercier, M.~Khabbazian, and V.~Bhargava, ``On the number of subsequences
  when deleting symbols from a string,'' \emph{{IEEE} Transactions on
  Information Theory}, vol.~54, no.~7, pp. 3279--3285, 2008.

\bibitem{swart_note_2003}
T.~G. Swart and H.~C. Ferreira, ``A note on double insertion/deletion
  correcting codes,'' \emph{{IEEE} Transactions on Information Theory},
  vol.~49, no.~1, p. 269–273, 2003.

\bibitem{tenengolts_nonbinary_1984}
G.~Tenengolts, ``Nonbinary codes, correcting single deletion or insertion
  (corresp.),'' \emph{{IEEE} Transactions on Information Theory}, vol.~30,
  no.~5, p. 766–769, 1984.

\bibitem{varshamov_arithmetic_1965}
R.~Varshamov, ``On an arithmetic function with an application in the theory of
  coding,'' \emph{Doklady Akademii nauk {SSSR}}, vol. 161, p. 540–543, 1965.

\bibitem{varshamov_codes_1965}
R.~Varshamov and G.~Tenengolts, ``Codes which correct single asymmetric
  errors,'' \emph{Avtomatika i Telemekhanika}, vol.~26, p. 288–292, 1965.

\end{thebibliography}

\end{document}